\documentclass[runningheads]{llncs}
\usepackage{amsmath}
\usepackage[version=4]{mhchem}
\usepackage{mathtools}
\usepackage{amssymb}
\usepackage{algorithm}
\usepackage[noend]{algpseudocode}
\usepackage{commath}
\usepackage{xcolor}
\usepackage{graphicx}
\graphicspath{{Graphics/}}
\usepackage{hyperref}

\usepackage{tabularx,booktabs}
\newcolumntype{Y}{>{\centering\arraybackslash}X}

\begin{document}
\title{Bayesian learning of effective chemical master equations in crowded intracellular conditions\thanks{Supported by Leverhulme Trust.}}

\titlerunning{Learning crowded effective CME}
%
%
\author{Svitlana Braichenko\inst{1,2}\orcidID{0000-0003-3330-6631} \and
Ramon Grima\inst{2}\orcidID{0000-0002-1266-8169} \and
Guido Sanguinetti\inst{1,3}\orcidID{0000-0002-6663-8336}}
\authorrunning{S. Braichenko et al.}
%
\institute{School of Informatics, \and
School of Biological Sciences, University of Edinburgh, Edinburgh, UK
\\
\and
SISSA, Trieste, Italy}
%

\maketitle              
\begin{abstract}
	Biochemical reactions inside living cells often occur in the presence of crowders - molecules that do not participate in the reactions but influence the reaction rates through excluded volume effects. However the standard approach to modelling stochastic intracellular reaction kinetics is based on the chemical master equation (CME) whose propensities are derived assuming no crowding effects. Here, we propose a machine learning strategy based on Bayesian Optimisation utilising synthetic data obtained from spatial cellular automata (CA) simulations (that explicitly model volume-exclusion effects) to learn effective propensity functions for CMEs. The predictions from a small CA training data set can then be extended to the whole range of parameter space describing physiologically relevant levels of crowding by means of Gaussian Process regression. We demonstrate the method on  an enzyme-catalyzed reaction and a genetic feedback loop, showing good agreement between the time-dependent distributions of molecule numbers predicted by the effective CME and CA simulations.

\end{abstract}

\section{Introduction}\label{sec:intro}
The empirical demonstration of stochasticity in gene expression \cite{elowitzStochasticGeneExpression2002} has had a profound impact both on experimental and computational biology. Experimentally, the last two decades have witnessed a flourishing of advanced technologies to measure stochastic effects in biology at unprecedented throughput and spatial/temporal resolution \cite{darzacq2009imaging,shah2018dynamics,larsson2019genomic}. Computationally, considerable effort has gone towards developing algorithmic solutions to facilitate the {\it in silico} simulation of stochastic biological systems, and their calibration to observational data \cite{gillespie1977exact,voliotis2016stochastic,gillespie2000chemical,schnoerrApproximationInferenceMethods2017,suter2011mammalian,skinner2016single}.

The vast majority of stochastic modelling work operates within the framework of the classical Chemical Master Equation (CME), which describes the time evolution of the (single-time marginal) state probability distribution of a discrete state, continuous-time Markovian system \cite{vankampenStochasticProcessesPhysics2007}. In this framework, each reaction has associated with it a propensity function which is derived assuming molecules are point particles diffusing fast enough such that 
well-mixed conditions ensue \cite{gillespie1992rigorous,vankampenStochasticProcessesPhysics2007,gillespie2009diffusional}. The CME formulation provides a number of advantages, including a transparent and elegant mathematical formalism, and efficient simulation and approximation algorithms \cite{schnoerrApproximationInferenceMethods2017}. In particular, the existence of an exact Stochastic Simulation Algorithm (SSA) \cite{gillespie1977exact} has led to the wide use of this formulation. However, the assumption that reactants freely diffuse inside cells is clearly at odds with biological reality: the cellular environment is spatially highly structured, and, for every given reaction, it contains large numbers of particles that do not partake in the reaction, creating a crowding effect which can significantly affect the dynamics of biochemical processes inside the cell \cite{van2000macromolecular,zhou2008macromolecular,tan2013molecular,mourao2014connecting}. While the modelling community is keenly aware of this mismatch in assumptions, there have been only a few attempts at modifying the propensities of the SSA/CME to take into account crowding \cite{grima2010intrinsic,cianci2016molecular,gillespie2007effect}. These pioneering studies have focused on simple biochemical systems but they do not provide a general recipe applicable to all intracellular reaction systems of interest. In contrast, particle-based algorithms (such as Brownian dynamics and cellular automata \cite{berryMonteCarloSimulations2002,schnellReactionKineticsIntracellular2004,grimaSystematicInvestigationRate2006,grima2010intrinsic,smith2017fast,kim2010crowding,chew2018reaction,andrews2017smoldyn}) have been extensively used to study the effect of crowding; these approaches while naturally suited to study crowding, they are computationally demanding since they model the movement of each particle  (crowder or reactant) in the system. 

In this paper, we devise a computational approach based on machine learning to adapt the efficiency of the CME approach to the reality of crowding. In a nutshell, the idea is to learn CME propensity functions that lead to particle number distributions which optimally match the ones resulting from particle-based algorithms. This leads to a general computational recipe for constructing effective CMEs that capture the stochastic dynamics of any biochemical system in crowded conditions.  


\section{Connecting different mathematical descriptions of stochastic kinetics using Bayesian optimization}

In this paper, we shall be concerned with two different descriptions of stochastic chemical kinetics: (i) Cellular automata (CA); (ii) Monte Carlo simulations using the SSA. We next describe each in detail and then show how a Bayesian optimization based procedure can be used to connect these different stochastic descriptions.

\subsection{Cellular Automata}

\begin{figure}[h]
	\centering
	\resizebox{\textwidth}{!}{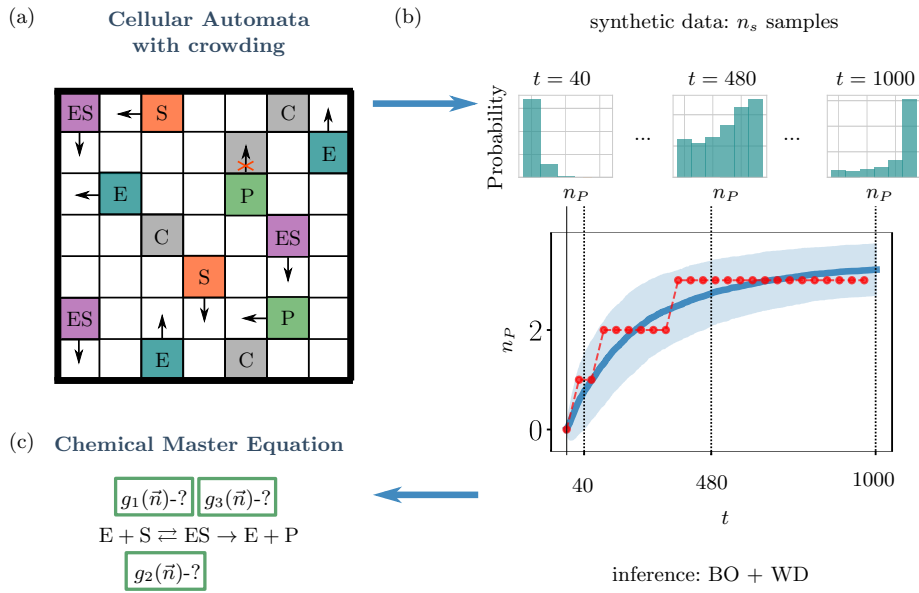}
	
	\caption{Cartoon illustrating the learning procedure, exemplified by means of a simple enzyme reaction. (a) First, we generate synthetic data using cellular automata simulations of the enzyme reaction. We show reactants (substrate S in orange and enzyme E in teal), enzyme-substrate complexes (ES in purple), products (P in green), inert crowders (C in grey) and empty spaces (white). The particle movements are updated according to a set of rules (see Appendix~\ref{CA}). The arrows show the particle movement directions; note that a particle can only move if an empty space is available. (b) We collect $n_s$ CA sample trajectories of the numbers of molecules of E, S, ES and P in time (a typical trajectory for the number of P molecules is shown as red connected dots while the mean and standard deviation over the trajectories are shown by a solid blue line and a shaded blue area, respectively). The marginal distributions of the number of each species molecules sampled at a number of discrete time points constitute our synthetic data (we show only those for P). (c) Finally we use Bayesian optimization (BO) to minimize the Wasserstein distance (WD) between the time-dependent particle number distributions generated by the CA and the SSA; this leads to the propensities $g_i(\vec{n})$ of the effective CME.}

	\label{fig:inf_cartoon}
\end{figure}

Cellular Automata (CA) can be characterized as a lattice of sites each holding a finite number of discrete states plus some rules which describe the evolution of the state of each site. Typically these update rules are a function of the states of the sites within a local neighbourhood. For a general introduction to CA in the context of biological and chemical modelling see \cite{deutsch2005mathematical,wolf2004lattice}. Many CA models that study the influence of crowding in biochemical reactions \cite{berryMonteCarloSimulations2002,schnellReactionKineticsIntracellular2004,grimaSystematicInvestigationRate2006} have the following properties in common: (i) each lattice site is either occupied by a molecule or empty; (ii) at each time step, a molecule is selected at random and one of its neighbouring sites is also chosen at random; (iii) if the chosen neighbouring site is empty then the molecule moves to it otherwise a reaction is attempted (only if the site is occupied by a reactant). In Fig. \ref{fig:inf_cartoon}(a) we illustrate a CA modelling a simple enzymatic reaction. 

The main advantage of CAs is their simplicity, in particular it is easy to devise rules that mimic molecular movement and interaction in complex geometries. Their disadvantages are (i) each molecule, independent of type, occupies the same amount of space (equal to one lattice site); (ii) the regularity of the lattice can influence the simulated dynamics. These main disadvantages can be overcome by using Brownian-dynamics, a lattice-free approach \cite{kim2010crowding,wieczorek2008influence,grima2010intrinsic,smith2017fast}, however these simulations are much more computationally expensive and we do not consider them further here.

\subsection{The chemical master equation and the SSA}
An alternative mathematical description to CA involves ignoring the spatial information and deriving equations that characterize the statistics of the total number of particles of each species in the volume of interest. Specifically, the system may be described in terms of the state vector $\mathbf{n}=(n_1, ... , n_N)$, where $n_i$ indicates a number of molecules belonging to species $X_i$. The dynamics of the reaction system can be described in terms of the probability distribution $P(\mathbf{n}, t) = P(\mathbf{n}, t | \mathbf{n_0}, t_0)$ for the system to be in state $\mathbf{n}$ at time $t$ when it was in state $\mathbf{n_0}$ at time $t_0$. The time evolution of this probability distribution obeys a master equation
\begin{equation}
\partial_t P(\mathbf{n}, t) = \sum_{r=1}^R g_r(\mathbf{n - S_r})P(\mathbf{n - S_r}, t) - \sum_{r=1}^R g_r(\mathbf{n})P(\mathbf{n}, t), \label{eq:CME}
\end{equation}
where $g_r(\mathbf{n}) \Delta t$ is the probability of reaction $r$ occurring somewhere in the compartment in the time interval $[t,t+\Delta t)$. The propensity functions $g_r(\mathbf{n})$ have been derived from first principles when the interacting particles are point-like (no volume exclusion) and assuming that all species are well-mixed, i.e., when the distance travelled by molecules between successive reactions is much larger than the size of the system \cite{gillespie1992rigorous,gillespie2009diffusional}. In this case, the master equation Eq. \eqref{eq:CME} is known as the chemical master equation. In the well-mixed limit, this is formally equivalent to the reaction-diffusion master equation \cite{gardinerStochasticMethodsHandbook2009} which has been shown to provide an accurate approximation of microscopic simulations that track point particle positions \cite{baras1996reaction}. We note that there is no general closed-form solution to the CME and hence in practice, one uses the SSA ~\cite{gillespieStochasticSimulationChemical2007} to estimate $P(\mathbf{n}, t)$ in a Monte Carlo setup. We refer the reader to \cite{gardinerStochasticMethodsHandbook2009,schnoerrApproximationInferenceMethods2017} for comprehensive background on the CME and the various methods to approximate its solutions.

\subsection{Bayesian Optimisation (BO)\label{ssec:BO}}
The CME presents considerable computational advantages over CA simulations, however simply ignoring spatial effects will generally result in an inaccurate prediction of the stochastic dynamics of the total number of particles for each species in a compartment. Nevertheless, it is plausible that there exists a different CME parametrisation which leads to a time-dependent probability distribution of molecule numbers that well approximates the same distributions calculated from CA. The main purpose of this paper is to illustrate a machine-learning strategy that automates the task of finding appropriate propensity functions. 


The task is akin to the inference of CME parameters from  observations \cite{schnoerrApproximationInferenceMethods2017,loskotComprehensiveReviewModels2019,ocalParameterEstimationBiochemical2019}, however in this case the CME parameters are not determined in order to maximise a data likelihood, but rather to give rise to (transient and steady state) particle number distributions that match the distributions generated from CA. 
To do so, we use a machine learning technique called Bayesian optimisation (BO)~\cite{shahriariTakingHumanOut2016,brochuTutorialBayesianOptimization2010}. BO is an efficient sequential algorithm to optimise objective functions which are very expensive to evaluate. In our case, the task is
\begin{equation}
\mathbf{x^*} = \underset{{\mathbf{x}}}{ \arg \min}f(\mathbf{x}),
\label{objfn}
\end{equation}
where $\mathbf{x}\in\mathbb{R}^d$ is a vector of CME parameters and the objective function $f(\mathbf{x})$ is a (rescaled) one-dimensional Wasserstein distance (WD)~\cite{villaniOptimalTransportOld2009} between the empirical marginal distribution of CA and CME trajectories \eqref{eq:CME}; see Appendix A for more details. BO works by using discrete evaluations of the expensive objective function $f(\mathbf{x})$ to construct a statistical surrogate (the {\it acquisition function}) which is easy to optimise and can be used to identify the next query point. Generally, Gaussian process (GP) regression  \cite{rasmussenGaussianProcessesMachine2006} is used to construct the surrogate function. GP regression is a methodology to perform Bayesian inference over functional spaces; given some training points (in our case estimates of the objective function at some parameter values), one can obtain a posterior predictive distribution over function values everywhere in parameter space, in terms of a mean posterior function $\mu(\mathbf{x})$ and posterior variance function $\sigma(\mathbf{x})$, which provide a point-wise posterior distribution over the values of the expensive function $f(\mathbf{x})$. The acquisition function is then obtained as some analytical function of the posterior mean and variance. In our case, we use the expected improvement acquisition function
	\begin{equation}
	    \alpha_{EI}(\mathbf{x})=(\mu_n(\mathbf{x}) - \tau_n)\Phi\left(\frac{\mu_n(\mathbf{x})- \tau_n}{\sigma_n(\mathbf{x})}\right)+\sigma_n(\mathbf{x})\phi\left(\frac{\mu_n(\mathbf{x})- \tau_n}{\sigma_n(\mathbf{x})}\right),
	\end{equation}
	where $\tau_n$ is the highest value of the unknown function found after $n$ steps, $\mu_n(\mathbf{x})$ and $\sigma_n(\mathbf{x})$ are the posterior mean and variance from GP regression after $n$ function evaluations, $\Phi$ is the standard normal cumulative distribution function, and $\phi$ is the standard normal probability density function. The acquisition function is analytically computable and can be optimised using standard methods to provide an optimal query point for the next evaluation of the expensive function $f(\mathbf{x})$. We note that our choice of acquisition function is not unique; other possible choices are upper confidence bound, probability of improvement and knowledge gradient \cite{shahriariTakingHumanOut2016}. The expected improvement acquisition function that we use is guaranteed to find the optimum of the target function under mild assumptions \cite{vazquez2010convergence}.
	
%
The computational recipe is as follows:

\begin{enumerate}
	\item Generate a number $n_s$ of CA trajectories for some chosen set of parameter values, and collect marginal probabilities of particle numbers at a set of time points.
	\item 
	Generate initial estimates of $f(\mathbf{x})$ by evaluating Wasserstein distances between CA distributions and CME distributions on a grid of CME parameters, and obtain the optimal initial value $\tau_0$ as the minimum distance obtained.
	\item Perform a sequential search which, at iteration $n$, selects a location $\mathbf{x}_{n}$ at which to query $f(\mathbf{x}_{n})$ by maximising the expected improvement acquisition function. Set $\tau_n=\mathrm{max}\{\tau_{n-1},f(\mathbf{x}_{n})\}$ and recompute the acquisition function by including $f(\mathbf{x}_{n})$ among the training points. 
	\item
	Once the improvement in objective function becomes smaller than a pre-defined threshold, the algorithm makes a final recommendation $\mathbf{x}_p$, which represents the best estimate for the minimum of $f(\mathbf{x})$.
	
\end{enumerate}
\begin{figure}[h]
	\centering
	\resizebox{\textwidth}{!}{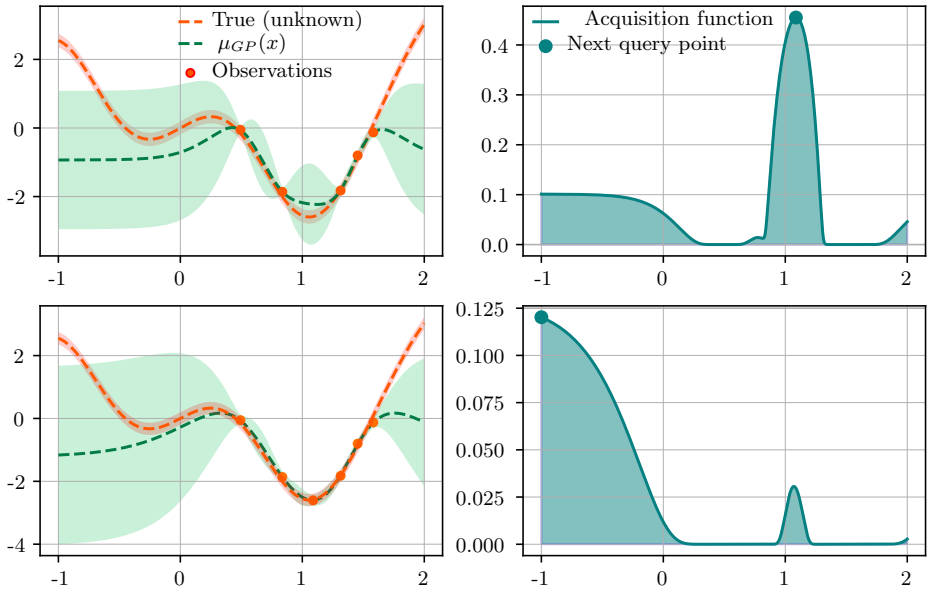}
	
	\caption{Illustration of how BO works to find the minimum of an arbitrary function. The GP prediction is characterized by its mean (dashed green line) and variance (shaded green area). The true function of interest is shown using a red dashed line with shading denoting an uncertainty range.  After observing the function for 5 times (red dots) -- see top left figure -- we select the next observation point by calculating and maximizing the acquisition function shown by the teal line -- see top right figure. The bottom row of plots shows how the prediction approaches the true function with more observations. In this paper, the BO procedure is used to find the reaction rate values in the propensity functions of the CME (the parameter vector $\mathbf{x}$) which minimize the mean WD (the function) between the time-dependent distributions computed by CA and the CME.}
	\label{GP_cartoon}
\end{figure}
We use the python BO implementation in scikit-optimize and in parallel we confirm the results using the python package optuna.

 A cartoon summarizing the overall learning procedure is shown in Fig.~\ref{fig:inf_cartoon} and an illustration focusing on how BO works to minimize a function is shown in Fig.~\ref{GP_cartoon}.
 
The BO procedure allows us to associate with one set of CA parameters a parametrisation of an effective CME which optimally matches (in a Wasserstein distance sense) the marginal probabilities of the CA. To extend this to {\it any} parametrisation of the CA, we once again appeal to smoothness and use CA/ CME pairs on a grid of CA parameters (each obtained via a separate BO procedure) to train a GP regression map to predict effective CME parametrisations also for unseen CA systems. This allows us to avoid expensive CA simulations and to provide effective CMEs for any setting of biochemical and crowding parameters in the CA system.

\section{Applications}\label{sec:res}

In this section, we apply our BO-based method to an enzyme system and a gene regulatory  network. The code to recreate the results as well as the CA data are available at \href{https://github.com/sb2g14/wasserstein\_time\_inference}{https://github.com/sb2g14/wasserstein\_time\_inference}.


\subsection{Michaelis-Menten reaction in crowded conditions \label{Enzyme}}


Here we study the stochastic kinetics of Michaelis-Menten enzyme reaction system
\begin{align} \label{eq:enz1}
&\text{E} + \text{S} \xrightleftharpoons[k_{-1}]{k_{1}}\text{ES} \xrightarrow{k_2} \text{E} + \text{P},
\end{align}
where $\text{E}$, $\text{S}$, $\text{ES}$ and $\text{P}$ are enzyme, substrate, enzyme-substrate complex and protein, respectively; $k_1$, $k_{-1}$ and $k_2$ are bimolecular, reverse and catalytic rates, respectively. Specifically, we consider this reaction in crowded conditions, where the (inert) crowders are assumed to be immobile ~\cite{berryMonteCarloSimulations2002,schnellReactionKineticsIntracellular2004}. The detailed set of rules for the CA simulations of this system is described in Appendix~\ref{CA}. 
   
Similar to~\cite{schnellReactionKineticsIntracellular2004}, the reaction rates may be estimated from the CA simulations directly using the formulae
\begin{align}
k_1 &= \frac{\dif \gamma / \dif t}{[E][S]}, \label{enz:k1}\\
k_{-1}&=\frac{\dif \gamma / \dif t + \dif \ [S] / \dif t}{[ES]}, \label{enz:km1}\\
 k_2 &= \frac{\dif \ [P]/ \dif t}{[ES]} \label{enz:k2},
\end{align}
where $\gamma(t)$ is the average number of $\text{E} + \text{S} \xrightarrow{}\text{ES}$ reactions that have occurred in the time interval $[0, t]$ divided by the number of grid points and $[X]$ is the concentration of species $X$, i.e. the average number of particles of species $X$ divided by the number of grid points. Note that averages are here understood to be computed over an ensemble of independent CA simulations. The reactions occur on a 2D square lattice of size $100 \times 100$. The initial values of the molecule numbers of each species are distributed uniformly and the boundary conditions are periodic. The latter conditions are used since they typically give rise to small finite-size
errors in Monte Carlo simulations \cite{allen2017computer}. 

\begin{figure}[h]
	\centering
	\includegraphics[width=\textwidth]{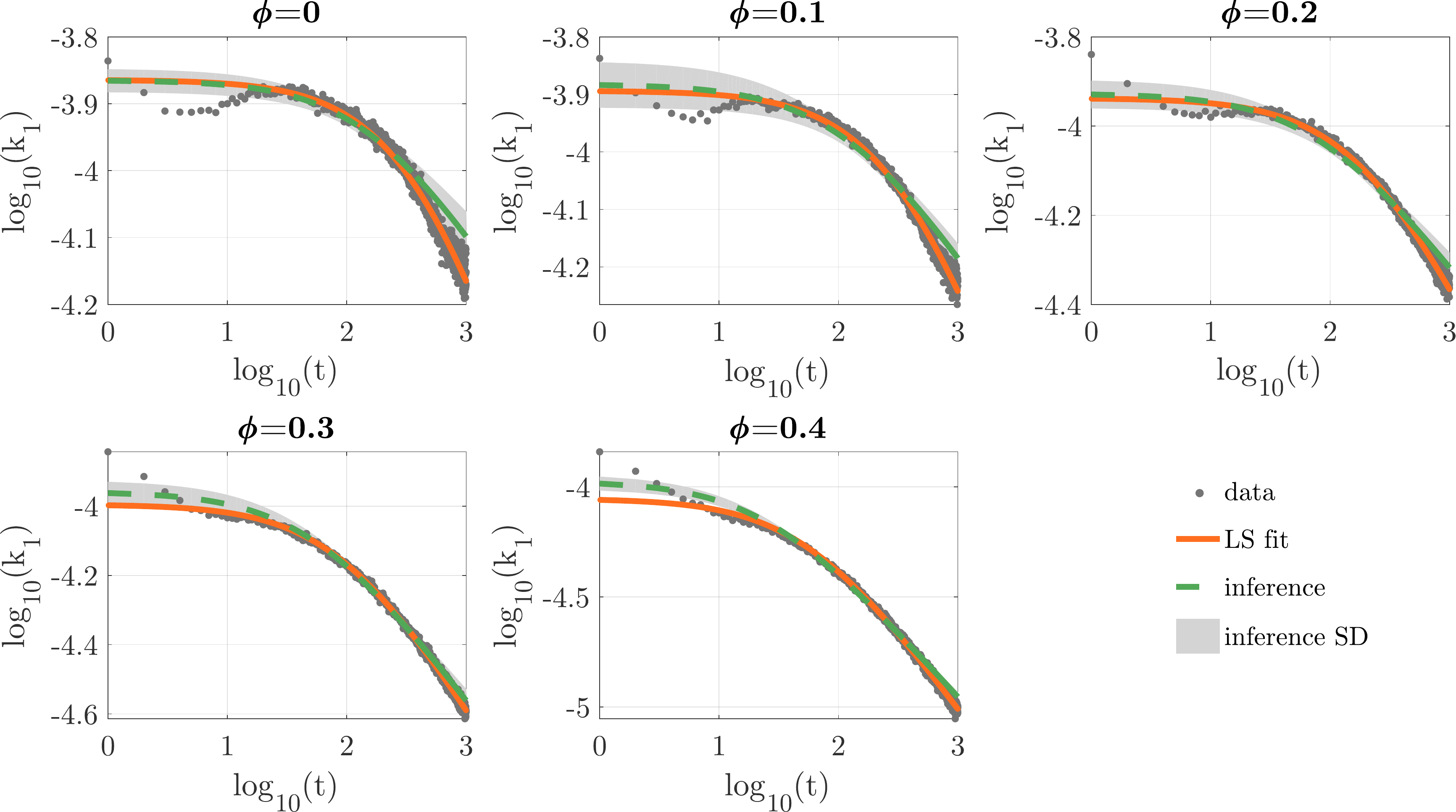}
	\caption{Variation of the bimolecular reaction rate with time and the concentration of crowders ($\phi$). The dark grey data points show the direct calculation of the rates using Eq.~\eqref{enz:k1} where the concentrations and $\gamma$ are calculated from an ensemble of $2000$ trajectories generated using CA simulations. The time-dependent rates are fit to the functions $k_1=\frac{k_0}{(t+\tau)^h}$ (following a Zipf-Mandelbrot law) using the method of least squares (red lines). They are also estimated using $10$ BO runs (green dashed lines shows the average while the grey shaded area shows the standard deviation). The initial values of the molecule numbers are $N_{S}=2000$, $N_{E}=100$, $N_C=0$ and $N_P=0$. A detailed description of the CA simulations can be found in Appendix B. Note that time is in arbitrary units.}
	\label{fig:rates_enz}
\end{figure}

 In Fig.~\ref{fig:rates_enz}, we show the calculation of the effective bimolecular rate $k_1$ using Eq.~\eqref{enz:k1} as a function of time and of the concentration of crowders $\phi$ in the system (dark grey points). Note that the effective rates of the other two reactions $k_{-1}$ and $k_2$ (estimated using Eq.~\eqref{enz:km1} and Eq.~\eqref{enz:k2}) do not show any appreciable variation with crowding levels and hence we do not discuss them any further (the values of the estimated rates are in agreement with the probabilities of the associated reactions in the CA). Clearly, crowding induces a bimolecular rate that is monotonically decreasing with time -- this is due to the increasing amounts of product (and the decreasing amounts of the substrate) which reduces the rate of encounter of substrate and enzyme molecules. 

We fit the time-dependent bimolecular rates using the Zipf-Mandelbrot law $k_1=\frac{k_0}{(t+\tau)^h}$ with parameters $k_0$, $\tau$ and $h$ obtained from the least-squares fit of the data estimated from Eqs.~\eqref{enz:k1} -- these are shown are orange lines in Fig.~\ref{fig:rates_enz}. Note that $\phi$ was limited to the range $0-0.4$ since this is the physiological range \cite{schnellReactionKineticsIntracellular2004}.

We next aim to learn the parameters $k_0$, $\tau$ and $h$ that characterise the effective bimolecular reaction rate using BO. Specifically, we use BO to fit the time-dependent distributions of all species calculated from the CA with those obtained from SSA simulations where the propensity functions in the CME description (Eq.~\eqref{eq:CME}) are:
\begin{align}
    g_1(\vec{n}) &= \frac{k_0}{(t+\tau)^h} n_S n_E,  \label{eq:prop} \\
    g_2(\vec{n}) &= k_{-1} n_{ES}, \\
    g_3(\vec{n}) &= k_{2} n_{ES},
\end{align}
where $n_X$ is the number of molecules of species $X$. Since some of the propensities have a time-dependent rate coefficient, the SSA simulations cannot be performed using the standard Gillespie algorithm; rather we use the exact Extrande algorithm \cite{voliotis2016stochastic}. The objective function minimized by BO (see Eq.~\eqref{objfn}) is given by
\begin{align}\label{eq:obj}
f &= \sum_{i=1}^{N_t} f_i, \notag \\ f_i &= \frac{|\gamma^{ref}_{i} - \gamma^{SSA}_{i}|}{\gamma^{SSA}_{i}} + \sum_{j=1}^{N} \frac{\text{WD}(P^{j}_i, Q^{j}_i)}{\langle  Q^{j}_i \rangle},
\end{align}
where $\gamma^{ref}_{i}$ and $\gamma^{SSA}_{i}$ correspond to the sample averaged counter of bimolecular reactions in the system at time interval $[0,t_i]$ in the CA and SSA simulations, respectively; $N_t$ is the number of time intervals; $P^{j}_i$ and  $Q^{j}_i$ are the CA and SSA marginal distributions at time $t_i$ for the number of molecules of species $j$, respectively; $\langle  Q^{j}_i \rangle$ is the mean of the SSA number distribution $Q^{j}_i$; $N$ is the total number of species. The first term in Eq.~\eqref{eq:obj} helps to avoid parameters indistinguishably in the unimolecular reaction rates $k_{-1}$ and $k_{2}$. 

We repeat the BO-based estimation multiple times leading to a set of Zipf-Mandelbrot law curves -- in Fig.~\ref{fig:rates_enz}, dashed green lines show the mean of these functions while the shaded areas show their standard deviation. These are in good agreement with the least square estimates (orange lines) calculated previously. In Tab.~\ref{tab:inf_results} (Appendix C) we show the inferred parameters and the objective function $f$ for two different initial conditions. 

\begin{figure}[h]
	\centering
	\includegraphics[width=\textwidth]{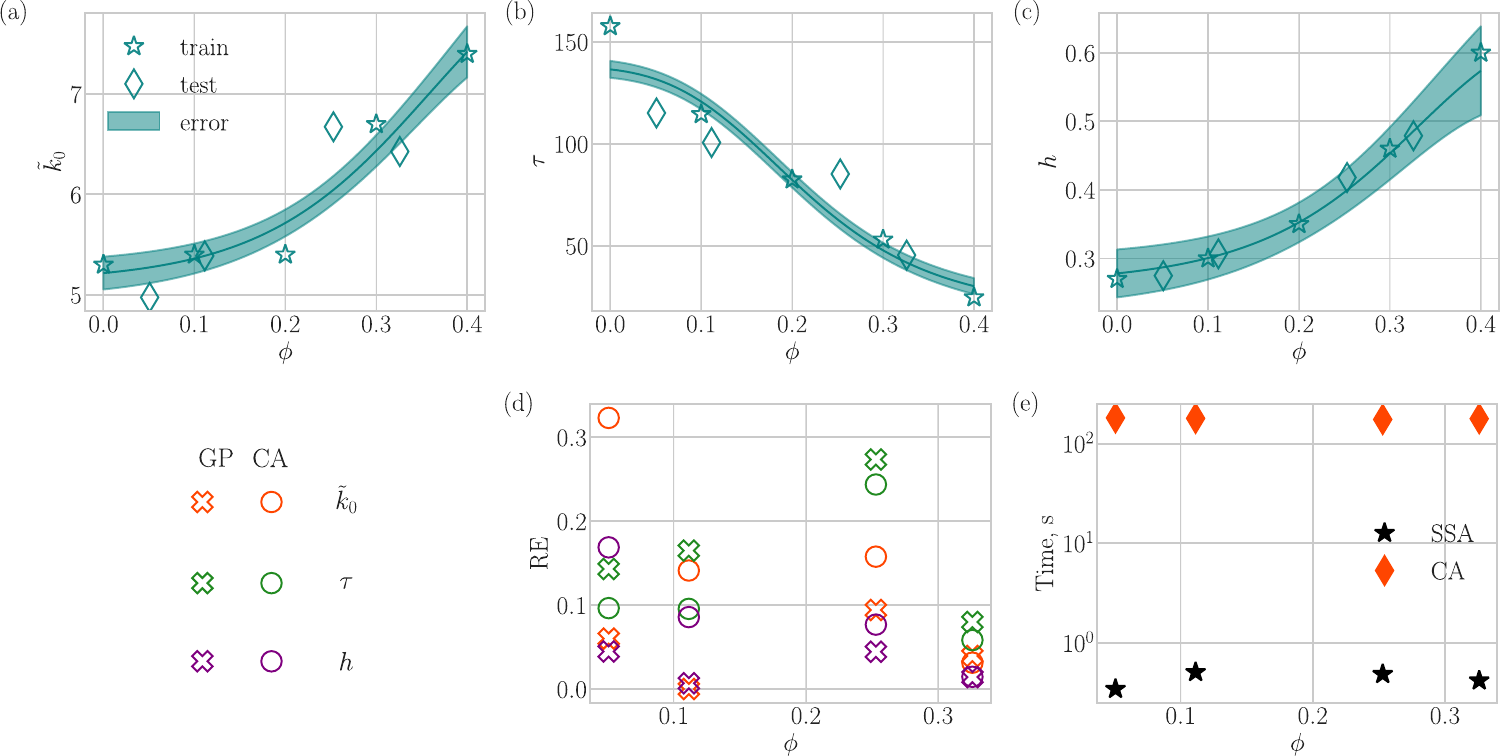}
	\caption{Learning the effective CME parameters for an enzyme system \eqref{eq:enz1} in crowded conditions. (a), (b) and (c) show the GP regression (blue line) of parameters of the Zipf-Mandelbrot law $\tilde{k}_0=k_0\Omega$ (here we rescale $k_0$ to avoid computational errors), $\tau$ and $h$, respectively; the prediction was built from 5 training points (teal stars) each obtained from BO trained with $n_s=2000$ CA simulation samples shown in Fig.~\ref{fig:rates_enz} and Tab.~\ref{tab:inf_results} (Appendix C). The shaded blue area shows the error in GP prediction. A number of test points (teal diamonds) each obtained from BO trained with $n_s=6000$ CA simulation samples fall on or close to the GP regression line showing its accuracy. (d) Relative testing errors in the effective rate as estimated indirectly by GP regression or directly by BO for 4 different testing points. Error is computed relative to the ground truth rates evaluated by BO from 6000 CA samples. (e) CPU runtimes of a single CA sample (teal diamonds) in comparison to training a single SSA sample (black stars) with predicted rates from pre-trained GP.}
	\label{fig:inf}
\end{figure}

To learn the functional dependence of the parameters of the effective bimolecular propensity $g_{1}(\vec{n})$ defined in Eq.~\eqref{eq:prop} with the crowding level $\phi$, we obtain effective parameters ($k_0,\tau,h$) by minimisation of the objective function~\eqref{eq:obj} for a set of training values of $\phi$ and then we use Gaussian Process (GP) regression to extend our predictions to a whole range of values of $\phi$ not covered by the training set. This approach has a huge computational benefit because we can obtain predictions of effective rates in the whole range of $\phi$ without running computationally demanding CA and BO at each point of the parameter space. Here we use the python GP implementation in scikit-learn with a sum of Neural Network kernel and white kernel, the latter term to model the noise induced by finite simulation samples. In Figs.~\ref{fig:inf}~(a), (b) and (c) we show the results of this procedure. The GP regression line (blue line) was here learnt from 5 training points (shown in teal stars and each obtained using BO trained with $n_s = 2000$ CA samples) evenly chosen in the space $\phi =0 - 0.4$. To test the accuracy of GP regression, we calculated $k_0$, $\tau$ and $h$ for another set of values of $\phi$; these testing points are shown as teal diamonds and are close to the GP regression curve calculated from training data.

To further test the accuracy of the GP regression line, we calculated the relative errors between the GP's estimates of $\tilde{k}_0=k_0\Omega$, $\tau$ and $h$ for 4 test points and a direct prediction from BO using 6000 CA samples at the same points (the ground truth). The results are shown by the open crosses in Fig.~\ref{fig:inf}~(d) -- the relative errors are relatively small showing the accuracy of the GP regression. We also compute the relative errors between the ground truth and the $\tilde{k}_0$, $\tau$ and $h$ directly predicted from BO using the same number of CA samples as used to train the GP. We see that this relative error (shown by the open circles) is of comparable magnitude to the one obtained earlier for the GP prediction; in other words, GP predictions appear to be of a similar quality to ab initio re-learning of the effective rates from a new batch of CA simulations. 

Finally, in Fig.~\ref{fig:inf}~(e) we show the time of training a new sample with SSA (black stars) given the pre-trained prediction model in comparison to running a single CA sample (red diamonds). Note that while we previously found that the errors of GP prediction are comparable to the CA errors, the training time for a new GP sample is more than $2$ orders of magnitude smaller without accounting for pre-training time. All the experiments were performed using a single core of Intel$^{\textregistered}$ Xeon$^{\textregistered}$ 3.5 GHz and 16 GB RAM.

\begin{figure}[h]
	\centering
	\includegraphics[width=\textwidth]{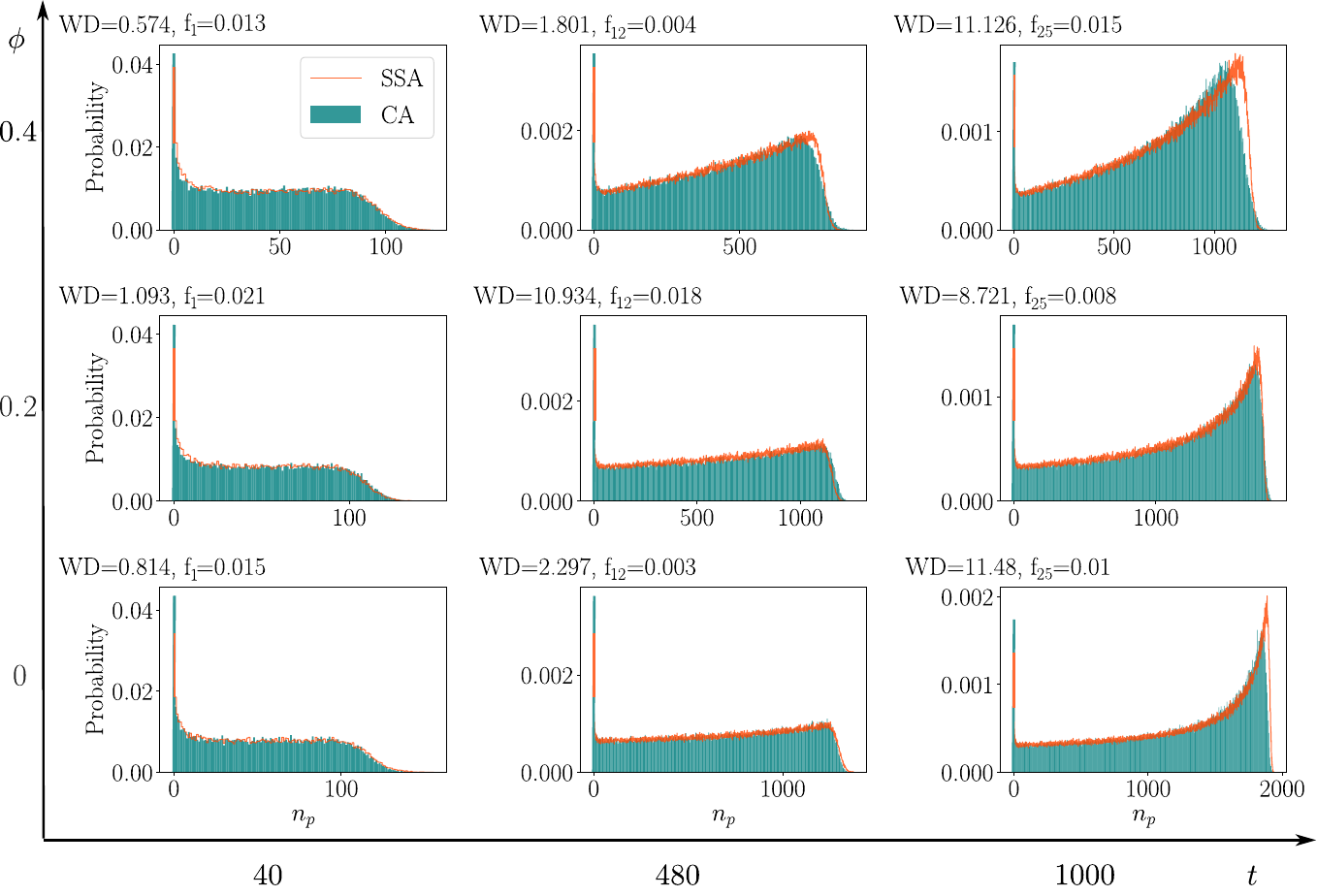}
	\caption{Comparing the marginal distributions of products of the enzyme reaction generated with the SSA using rates learned with BO (orange lines obtained with $n_s=500$) and data generated by CA (teal bars obtained with $n_s=2000$) at $\phi=0$, $\phi=0.2$ and $\phi=0.4$ for three different times $t=40,480,1000$. The WDs between both distributions as well as the value of $f_i$ (as given by Eq.~\eqref{eq:obj}) are shown in the corner of each subplot. Note that $f_i$ varies from $f_1$ to $f_{25}$ because we divide the period $[0,1000]$ into $N_t = 25$ subintervals.}
	\label{fig:cr_dist_CA}
\end{figure}

In Fig.~\ref{fig:cr_dist_CA} we compare distributions drawn from CA simulations of the enzyme reaction system~\eqref{eq:enz1} (our ground truth; teal histograms) and the distributions generated with SSA using effective rates calculated using BO (orange outline histograms). We compare the marginal distributions of products P obtained in the enzyme reaction. The left column in Fig.~\ref{fig:cr_dist_CA} compares distributions sampled at the beginning of the experiment ($t=40$) while the middle and the right columns compare the distributions in the middle ($t=480$) and the end ($t=1000$) of the experiment, respectively. 
The lower row of figures is for $\phi=0$, the middle row is for $\phi=0.2$ (middle row) and the top row is for $\phi=0.4$. In the corner of each subplot, we show the WD between the marginal distributions (obtained from the CA and SSA+BO) and the value of $f_i$ (as defined by Eq.~\eqref{eq:obj}) for the respective time interval. 
Curiously, the distribution at $\phi=0$ and $t=480$, and the one at $\phi=0.2$ and $t=480$ look similar in the plot, but the difference in the WDs is significant at $2.3$ vs $10.9$. This is due to a systematic discrepancy between the CA and the SSA marginal distributions in the upper subplot that is not easily visible by eye but can be picked by zooming into the subplots. Also, we can see that the quality of the approximation degrades with an increase in $\phi$ (larger values of WD). This might happen because with an increase in crowding, the dynamics of the system become non-Markovian (time between successive reaction events is not any more exponential).

%

\subsection{Gene network with negative feedback}
Next, we study the stochastic kinetics of a gene network with negative feedback in the presence of crowding (a well-mixed, non-crowded version of this system was studied in~\cite{thomasSlowscaleLinearNoise2012}). 

\begin{figure}[h]
	\centering
	\includegraphics[width=3in]{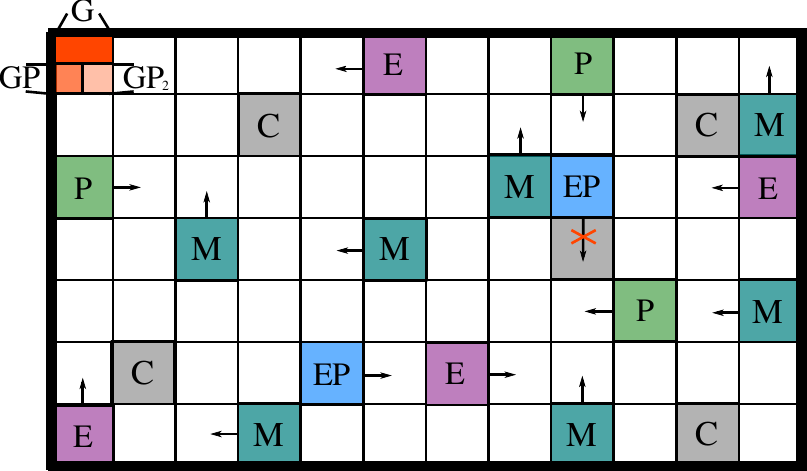}
	\caption{Cartoon illustrating the setup for the CA simulations of a genetic negative feedback loop inside a cell. The colours represent the following: G (dark orange), GP (orange) and GP$_2$ (light orange) are different states of the gene with different numbers of protein-bound 0, 1, 2, respectively; protein P (green), mRNA M (teal), enzyme E (purple), complex EP (blue) and crowders C (grey). White represents empty space. All molecules can move except crowders and the gene which are immobile. The arrows show possible directions of movement of the molecules -- movement is only possible if a neighbouring space is empty thus enforcing volume exclusion. For the possible reactions between molecules see the main text.}
	\label{fig:gene_cartoon}
\end{figure}

The CA simulations for this system proceed via a set of rules and boundary conditions, akin to those used previously for enzyme kinetics. We consider a fictitious 2D cell defined by a square lattice of points ($100 \times 50$) with periodic boundary conditions (to reduce finite-size effects). One (immobile) lattice point is the gene which can have one of three states: G (unbound to a protein), GP (bound to a protein) and $\rm{GP}_2$ (bound to two proteins). The rest of the lattice points are either empty or occupied by an mRNA (M), a protein (P), a crowder (C), a free degrading enzyme (E) or a protein-enzyme complex (EP). All of these molecules are mobile, i.e. can jump to a neighbouring empty lattice point, except the crowders which are immobile at all times. M is produced when the gene is in state G or GP (transcription); subsequently, M can produce P (translation) or else it is removed from the system (mRNA degradation). P can bind to the enzyme E to form the complex EP which can then decay to E (protein degradation). P can also bind to the gene G to form GP and this can bind to another P to form $\rm{GP}_2$. G and GP are assumed to produce M at the same rate but $\rm{GP}_2$ is transcriptionally silent -- hence this is a negative feedback loop since the gene-product (the protein) represses its own production. In all cases, the initial number of molecules of each species are $n_G=1$, $n_{GP} = 0$, $n_{GP2} = 0$, $n_M = 0$, $n_P = 0$, $n_E=100$, $n_{EP}=0$. In Fig.~\ref{fig:gene_cartoon} we show a cartoon of this system. 

The procedure leading to an effective CME, approximating the spatial CA dynamics, is the same as before. We use BO to fit CA generated time-dependent marginal distributions of all species to those generated by an SSA. In this case the reactions modelled by the SSA are given by
\begin{align} \label{eq:GN}
&\text{G} \xrightarrow{k_0} \text{G} + \text{M}, \quad \text{M} \xrightarrow{k_s} \text{P} + \text{M}, \quad \text{P} + \text{E} \xrightleftharpoons[k_{-3}]{k_{3}}\text{EP} \xrightarrow{k_4} \text{E}, \quad \text{M} \xrightarrow{k_{dM}} \varnothing, \nonumber \\ &\text{P} + \text{G} \xrightleftharpoons[k_{-1}]{k_{1}}\text{GP}, \quad \text{P} + \text{GP} \xrightleftharpoons[k_{-2}]{k_{2}}\text{GP}_2, \quad \text{GP} \xrightarrow{k_0}\text{GP} +\text{M}.
\end{align}
Note that the objective function minimized by BO is same as Eq.~ \eqref{eq:obj} but without the (first) $\gamma$ dependent term; to lighten the computational burden, we choose to infer only the three bimolecular rates ($k_1$, $k_2$, $k_3$) and assume that the unimolecular reaction rates are fixed to the ones set in the CA simulations. This is a reasonable assumption since crowding tends to primarily affect bimolecular rates. Note that while the inferred bimolecular rates were time-dependent in the previous enzyme example, for the feedback loop they are found to quickly converge to a time-independent non-zero value and hence we do not need to assume a Zipf-Mandelbrot law for the rates -- this is because for the feedback loop, in steady-state all species numbers fluctuate around a non-zero value. 

\begin{figure}[h]
	\centering
	\includegraphics[width=\textwidth]{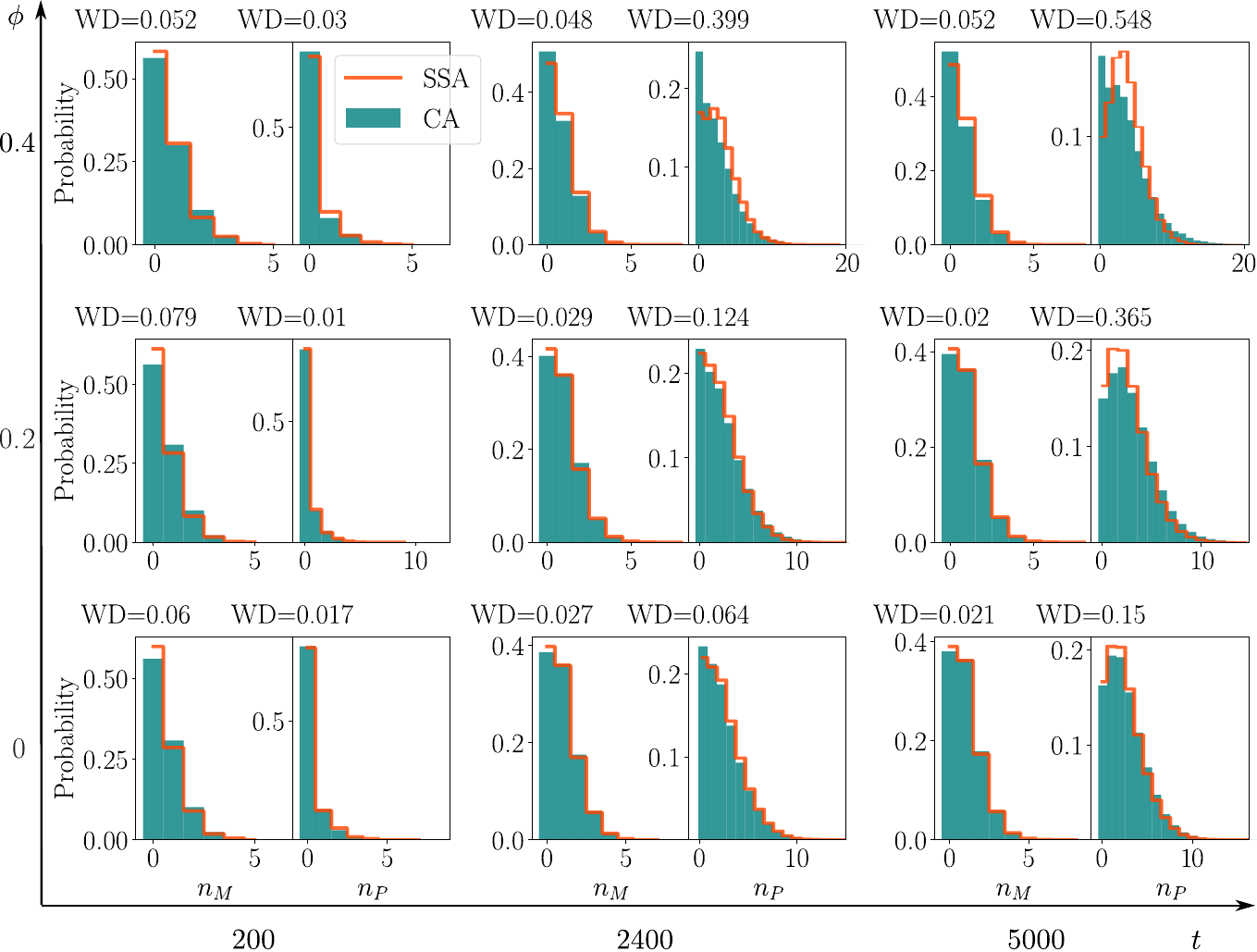}
	\caption{Comparing the marginal distributions of mRNA and protein in a genetic feedback loop generated with the SSA using rates learned with BO (orange lines obtained with $n_s=500$) and data generated by CA (teal bars obtained with $n_s=2000$) at $\phi=0$, $\phi=0.2$ and $\phi=0.4$ for three different times $t=200,2400,5000$. The bimolecular reaction rates are inferred while the unimolecular rates are fixed at the same values as the CA (see Tab.~\ref{tab:gene_rates} in Appendix C). The WDs between the distributions are shown on the top of each subplot.}
	\label{fig:gene_distributions_M_P}
\end{figure}

\begin{figure}[h]
	\centering
	\includegraphics[width=\textwidth]{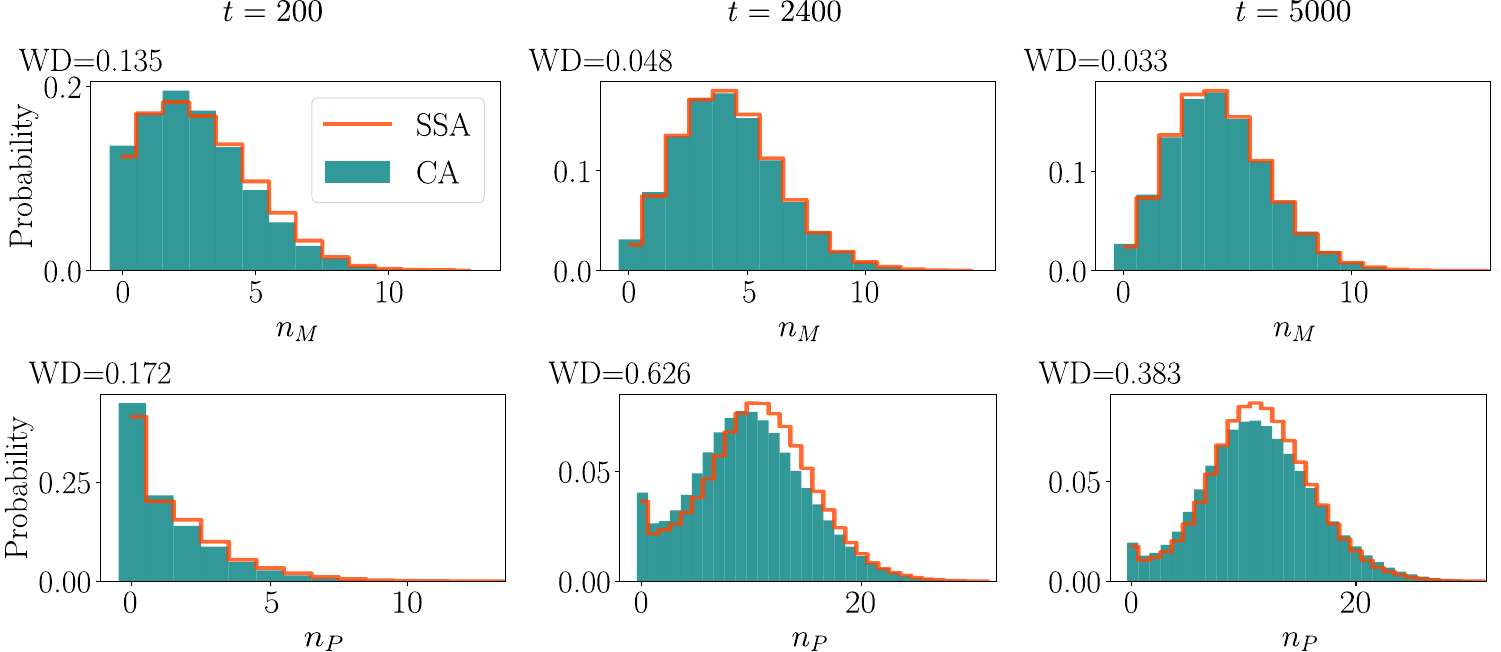}
	\caption{
	Comparing the marginal distributions of mRNA and protein in a genetic feedback loop generated with the SSA using rates learned with BO (orange lines obtained with $n_s=500$) and data generated by CA (teal bars obtained with $n_s=2000$) at $\phi=0$ for three different times $t=200,2400,5000$. The inferred bimolecular reaction rates are $k_1 = 0.0018$, $k_2 = 0.0025$, $k_3 = 0.0002$, while the unimolecular rates are fixed; the latter have the same values as in Tab.~\ref{tab:gene_rates}) (Appendix C) except that $k_0 = 0.05$.}
	\label{fig:gene_distributions_nonfit}
\end{figure}

The results of the parameter inference averaged over 5 BO runs are shown in Tab.~\ref{tab:gene_rates} in Appendix C (here we also show the probabilities of the individual reactions in the CA). The standard deviation in the inferred parameters in most of the cases is smaller than $20\%$ of the mean, therefore, we can conclude that the parameters are inferred fairly well. In Fig. \ref{fig:gene_distributions_M_P} we compare distributions drawn from CA simulations (our ground truth; teal histograms) and the distributions generated with SSA using effective rates calculated using BO (orange outline
histograms). Note the same tendency as in the previous example, where the WDs increase with the level of crowding probably because of the breakdown of the Markovian assumption behind the CME. 
Interestingly, as shown in Fig.~\ref{fig:gene_distributions_nonfit}, the CME starts to fail as a good approximation of the CA for increased mRNA production rate $k_0$ even in the case where there is no crowding. Presumably, this happens because of increased mRNA production close to the gene which causes a large degree of volume exclusion due to self-crowding of mRNA molecules (the CA sample average of the fraction of occupied volume in steady-state is quite low at $0.023$).

\section{Conclusions}
As advances in measurement technology probe deeper into the spatial stochasticity of biochemical reactions, novel computational tools are needed to formulate quantitative theories of cellular function. Existing frameworks for modelling spatial stochastic effects such as Brownian Dynamics simulations and Cellular Automata (CA) inescapably suffer from a high computational load. This is further aggravated by the frequent need to explore a range of parametrisations for the models as biochemical parameters are seldom accessible, creating the need for even larger-scale simulation studies. In this paper, we propose an automatic approach to generate simpler effective CME models which can recapitulate the statistical behaviour of spatially crowded stochastic systems. Given a (limited) number of expensive spatial simulation runs, our approach can provide a fast CME-based simulator for {\it any} parametrisation of the spatial system which optimally matches its statistical properties. Our approach focussed on CA spatial systems, but in principle, the same procedure can be deployed for any spatial simulator.

As well as providing an efficient simulation tool, our approach opens potential new directions. As a first application, its computational efficiency would easily allow the analysis of 3D systems, as the scaling of the CME is clearly independent of the dimension of the space in which the reactions happen. Secondly, the availability of efficient simulation tools opens the way to the use of simulator-based inference tools to estimate the parameters of spatial crowded systems from data \cite{cranmer2020frontier}, therefore enabling a formal statistical link between computational methodology and experimental technology.

\section*{Appendix}
\appendix

\section{Wasserstein Distance (WD)\label{WD}}
In principle any distance measure between distributions may be used as an objective function but WD has proven to be one of the most effective ~\cite{ocalParameterEstimationBiochemical2019}. Consider two distributions $P$ and $Q$ of datasets $P_1, ..., P_n$ and $Q_1, ..., Q_n$, then the Wasserstein distance between them is
\begin{equation}
\text{WD}^{(p)} = \left(\sum_{i=1}^{n}||P_i - Q_i||^p\right)^{\frac{1}{p}},
\end{equation}
where $p \ge 1$ is the dimensionality of the original data distribution. In this paper, we always use $p=1$.

\section{CA rules modelling enzyme kinetics in crowded conditions \label{CA}}

At the beginning of each simulation, the counter $\gamma$ is reset to zero, and E, S and crowder molecules are randomly placed on the square lattice. At each time step, a ``subject” molecule is randomly chosen and it is moved or participates in a reaction according to the following rules:

\begin{enumerate}

\item  Choose randomly one of 4 nearest neighbouring ``destination" sites. 
\item If the destination site is empty and the ``subject" molecule is E, S or P then move the molecule (simulates diffusion). 
\item Otherwise: 
	\begin{enumerate}
	\item If the ``subject" molecule is E or S and the molecule occupying the ``destination" site (``target" molecule) is, respectively, S or E then generate a uniform random number between 0 and 1. If this is lower than the reaction probability $P_1=1$, replace the ``target" molecule with ES, remove ``subject" molecule and increase the counter $\gamma$ by one. This step models the reaction $\text{E} + \text{S} \rightarrow \text{ES}$.
	\item If the ``subject" molecule is ES, check if there are any molecules placed on the neighbouring sites. If at least one nearest neighbour site is empty, randomly choose a vacant ``destination" site and generate a uniform random number between 0 and 1.
		\begin{enumerate}
		\item If the generated number is less than $P_{-1}=0.02$, place E on ``subject" site and S on ``destination" site ($\text{ES} \rightarrow \text{E} + \text{S}$).
		\item If the number is greater than $P_{-1}$ but lower than $P_{-1} + P_{2}$, where $P_2=0.04$, then place E on the ``subject" site and P on the ``destination" site ($\text{ES} \rightarrow \text{E} + \text{P}$).
		\item If the number is greater than $P_{-1} + P_{2}$ move ES to the ``destination" site (only diffusion occurs). 
		\end{enumerate}
	\end{enumerate}
\item Otherwise, if the ``destination" site is occupied, reject the move (simulates volume exclusion effects).

\end{enumerate}

\newpage 

\section{Supplementary tables}

\begin{table}[h]
	\centering
	\begin{tabular}{|c|c|c|c|c|c|}
		\hline
		Parameter & $\phi=0$ & $\phi=0.1$ & $\phi=0.2$ & $\phi=0.3$ & $\phi=0.4$ \\ \hline
		&  \multicolumn{5}{c|}{\bfseries BO $t_{max}=1000$, $N_{S}=2000$, $N_{E}=100$, $n_s=2000$,} \\
		&  \multicolumn{5}{c|}{ estimated from the mean over $10$ runs } \\
		\hline
		$k_0$ & $5.3 \cdot 10^{-4}$ &  $5.4 \cdot 10^{-4}$ &  $5.4 \cdot 10^{-4}$ &  $6.7 \cdot 10^{-4}$ &  $7.4 \cdot 10^{-4}$ \\
		$\tau$ & $157.6$ & $114.5$ & $82.5$ & $53.0$ & $24.7$ \\
		$h$ & $0.27$ & $0.3$ & $0.35$ & $0.46$ & $0.6$ \\
		$k_{-1}$ & $1.9 \cdot 10^{-2}$ & $2.0 \cdot 10^{-2}$  & $1.9 \cdot 10^{-2}$ & $1.9 \cdot 10^{-2}$ & $1.9 \cdot 10^{-2}$\\ 
		$k_2$ & $4.0 \cdot 10^{-2}$ & $4.0 \cdot 10^{-2}$ & $3.9 \cdot 10^{-2}$  & $3.9 \cdot 10^{-2}$ & $3.7 \cdot 10^{-2}$ \\ \hline
		$f$ & 1.6 & 1.5 & 1.4 & 1.4 & 1.6\\ \hline
		&  \multicolumn{5}{c|}{\bfseries BO $t_{max}=4000$, $N_{S}=3000$, $N_{E}=20$, $n_s=4000$, } \\ 
		&  \multicolumn{5}{c|}{ estimated from the mean over $10$ runs } \\ \hline
		$k_0$ & $8.2 \cdot 10^{-4}$ &  $8.6 \cdot 10^{-4}$ &  $5.5 \cdot 10^{-4}$ &  $1.0 \cdot 10^{-3}$ &  $1.0 \cdot 10^{-3}$ \\
		$\tau$ & $585.9$ & $551.2$ & $236.4$ & $248.7$ & $62.0$ \\
		$h$ & $0.27$ & $0.3$ & $0.29$ & $0.48$ & $0.67$ \\
		$k_{-1}$ & $2.0 \cdot 10^{-2}$ & $2.0 \cdot 10^{-2}$  & $1.9 \cdot 10^{-2}$ & $1.8 \cdot 10^{-2}$ & $1.7 \cdot 10^{-2}$\\ 
		$k_2$ & $4.0 \cdot 10^{-2}$ & $3.9 \cdot 10^{-2}$ & $3.8 \cdot 10^{-2}$  & $3.7 \cdot 10^{-2}$ & $3.4 \cdot 10^{-2}$ \\ \hline
		$f$ & 1.4 & 1.4 & 1.8 & 1.7 & 1.6\\ \hline
	\end{tabular}
	\vspace*{5mm}
	\caption{ Parameters $k_0$, $\tau$ and $h$ are estimated from the mean prediction of $k_1$ (as a function of time) over $6$ or $10$ BO runs for different $\phi$ in the CA simulations; $k_{-1}$, $k_2$ and $f$ are mean predictions. The estimates are for two different initial conditions; the top table is for the data shown in Fig.~\ref{fig:rates_enz}. \label{tab:inf_results}} 
\end{table}

\begin{table}[H]
	\centering
	\begin{tabular}{|c|c|c|c|c|c|}
		\hline
		Para- & $\phi=0$ & $\phi=0.1$ & $\phi=0.2$ & $\phi=0.3$ & $\phi=0.4$ \\ 
		meter &  &  &  &  &  \\
		\hline
		&  \multicolumn{5}{c|}{\bfseries BO $t_{max}=5000$, $n_s=2000$,} \\
		&  \multicolumn{5}{c|}{ estimated mean and standard deviation over $5$ runs } \\
		\hline
		$k_1 \cdot 10^{-4}$ & $17.7 \pm 3.0$ &  $23.3 \pm 3.4$ &  $31.8 \pm 3.1$ &  $44.6 \pm 5.1$ &  $66.8 \pm 3.8$ \\
		$k_2 \cdot 10^{-4}$ & $15.9 \pm 1.5$ &  $22.6 \pm 3.8$ &  $32.3 \pm 3.2$ &  $50.4 \pm 5.6$ &  $97.8 \pm 3.8$ \\
		$k_3 \cdot 10^{-5}$ & $21.3 \pm 2.4$ &  $21.5 \pm 1.6$ &  $20.6 \pm 0.7$ &  $16.4 \pm 0.6$ &  $11.5 \pm 0.4$ \\ \hline
	\end{tabular}
	\vspace*{5mm}
	\caption{ Parameters $k_1$, $k_2$ and $k_3$ are estimated from the mean predictions and standard deviation over $5$ BO runs for different $\phi$ in the CA simulations, unimolecular rate are fixed at $k_0 = 0.01$, $k_s = 0.01$, $k_{-3} = 0.05$, $k_4 = 0.01$, $k_{dM} = 0.01$, $k_{-1} = 0.02$, $k_{-2} = 0.05$. Note that the probability of each first-order reaction in the CA is the same as the corresponding rate constant above; the probability of each bimolecular reaction is set to 1 for simplicity. \label{tab:gene_rates}} 
\end{table}


\begin{thebibliography}{10}
	\providecommand{\url}[1]{\texttt{#1}}
	\providecommand{\urlprefix}{URL }
	\providecommand{\doi}[1]{https://doi.org/#1}
	
	\bibitem{elowitzStochasticGeneExpression2002}
	Elowitz, M.B., Levine, A.J., Siggia, E.D., Swain, P.S.: Stochastic {{gene
			expression}} in a {{single cell}}. Science  \textbf{297}(5584),  1183--1186
	(2002)
	
	\bibitem{darzacq2009imaging}
	Darzacq, X., Yao, J., Larson, D.R., Causse, S.Z., Bosanac, L., De~Turris, V.,
	Ruda, V.M., Lionnet, T., Zenklusen, D., Guglielmi, B., et~al.: Imaging
	transcription in living cells. Annual Review of Biophysics  \textbf{38},
	173--196 (2009)
	
	\bibitem{shah2018dynamics}
	Shah, S., Takei, Y., ~, W., Lubeck, E., Yun, J., Eng, C.H.L., Koulena, N.,
	Cronin, C., Karp, C., Liaw, E.J., et~al.: Dynamics and spatial genomics of
	the nascent transcriptome by intron seqfish. Cell  \textbf{174}(2),  363--376
	(2018)
	
	\bibitem{larsson2019genomic}
	Larsson, A.J., Johnsson, P., Hagemann-Jensen, M., Hartmanis, L., Faridani,
	O.R., Reinius, B., Segerstolpe, {\AA}., Rivera, C.M., Ren, B., Sandberg, R.:
	Genomic encoding of transcriptional burst kinetics. Nature
	\textbf{565}(7738),  251--254 (2019)
	
	\bibitem{gillespie1977exact}
	Gillespie, D.T.: Exact stochastic simulation of coupled chemical reactions. The
	journal of Physical Chemistry  \textbf{81}(25),  2340--2361 (1977)
	
	\bibitem{voliotis2016stochastic}
	Voliotis, M., Thomas, P., Grima, R., Bowsher, C.G.: Stochastic simulation of
	biomolecular networks in dynamic environments. PLoS Computational Biology
	\textbf{12}(6),  e1004923 (2016)
	
	\bibitem{gillespie2000chemical}
	Gillespie, D.T.: The chemical langevin equation. The Journal of Chemical
	Physics  \textbf{113}(1),  297--306 (2000)
	
	\bibitem{schnoerrApproximationInferenceMethods2017}
	Schnoerr, D., Sanguinetti, G., Grima, R.: Approximation and inference methods
	for stochastic biochemical kinetics\textemdash a tutorial review. Journal of
	Physics A: Mathematical and Theoretical  \textbf{50}(9),  093001 (2017)
	
	\bibitem{suter2011mammalian}
	Suter, D.M., Molina, N., Gatfield, D., Schneider, K., Schibler, U., Naef, F.:
	Mammalian genes are transcribed with widely different bursting kinetics.
	Science  \textbf{332}(6028),  472--474 (2011)
	
	\bibitem{skinner2016single}
	Skinner, S.O., Xu, H., Nagarkar-Jaiswal, S., Freire, P.R., Zwaka, T.P.,
	Golding, I.: Single-cell analysis of transcription kinetics across the cell
	cycle. Elife  \textbf{5},  e12175 (2016)
	
	\bibitem{vankampenStochasticProcessesPhysics2007}
	Van~Kampen, N.: Stochastic {{Processes}} in {{Physics}} and {{Chemistry}}.
	North-{{Holland}} Personal Library, {Elsevier}, {Amsterdam}, 3rd ed edn.
	(2007)
	
	\bibitem{gillespie1992rigorous}
	Gillespie, D.T.: A rigorous derivation of the chemical master equation. Physica
	A: Statistical Mechanics and its Applications  \textbf{188}(1-3),  404--425
	(1992)
	
	\bibitem{gillespie2009diffusional}
	Gillespie, D.T.: A diffusional bimolecular propensity function. The Journal of
	Chemical Physics  \textbf{131}(16),  164109 (2009)
	
	\bibitem{van2000macromolecular}
	Van~den Berg, B., Wain, R., Dobson, C.M., Ellis, R.J.: Macromolecular crowding
	perturbs protein refolding kinetics: implications for folding inside the
	cell. The EMBO journal  \textbf{19}(15),  3870--3875 (2000)
	
	\bibitem{zhou2008macromolecular}
	Zhou, H.X., Rivas, G., Minton, A.P.: Macromolecular crowding and confinement:
	biochemical, biophysical, and potential physiological consequences. Annual
	Review of Biophysics  \textbf{37},  375--397 (2008)
	
	\bibitem{tan2013molecular}
	Tan, C., Saurabh, S., Bruchez, M.P., Schwartz, R., LeDuc, P.: Molecular
	crowding shapes gene expression in synthetic cellular nanosystems. Nature
	Nanotechnology  \textbf{8}(8),  602--608 (2013)
	
	\bibitem{mourao2014connecting}
	Mour{\~a}o, M.A., Hakim, J.B., Schnell, S.: Connecting the dots: the effects of
	macromolecular crowding on cell physiology. Biophysical Journal
	\textbf{107}(12),  2761--2766 (2014)
	
	\bibitem{grima2010intrinsic}
	Grima, R.: Intrinsic biochemical noise in crowded intracellular conditions. The
	Journal of Chemical Physics  \textbf{132}(18),  05B604 (2010)
	
	\bibitem{cianci2016molecular}
	Cianci, C., Smith, S., Grima, R.: Molecular finite-size effects in stochastic
	models of equilibrium chemical systems. The Journal of Chemical Physics
	\textbf{144}(8),  084101 (2016)
	
	\bibitem{gillespie2007effect}
	Gillespie, D.T., Lampoudi, S., Petzold, L.R.: Effect of reactant size on
	discrete stochastic chemical kinetics. The Journal of Chemical Physics
	\textbf{126}(3),  034302 (2007)
	
	\bibitem{berryMonteCarloSimulations2002}
	Berry, H.: Monte {{Carlo simulations}} of {{enzyme reactions}} in {{two
			dimensions}}: {{fractal kinetics}} and {{spatial segregation}}. Biophysical
	Journal  \textbf{83}(4),  1891--1901 (2002)
	
	\bibitem{schnellReactionKineticsIntracellular2004}
	Schnell, S., Turner, T.E.: Reaction kinetics in intracellular environments with
	macromolecular crowding: Simulations and rate laws. Progress in Biophysics
	and Molecular Biology  \textbf{85}(2),  235--260 (Jun 2004)
	
	\bibitem{grimaSystematicInvestigationRate2006}
	Grima, R., Schnell, S.: A systematic investigation of the rate laws valid in
	intracellular environments. Biophysical Chemistry  \textbf{124}(1),  1--10
	(2006)
	
	\bibitem{smith2017fast}
	Smith, S., Grima, R.: Fast simulation of brownian dynamics in a crowded
	environment. The Journal of Chemical Physics  \textbf{146}(2),  024105 (2017)
	
	\bibitem{kim2010crowding}
	Kim, J.S., Yethiraj, A.: Crowding effects on association reactions at
	membranes. Biophysical Journal  \textbf{98}(6),  951--958 (2010)
	
	\bibitem{chew2018reaction}
	Chew, W.X., Kaizu, K., Watabe, M., Muniandy, S.V., Takahashi, K., Arjunan,
	S.N.: Reaction-diffusion kinetics on lattice at the microscopic scale.
	Physical Review E  \textbf{98}(3),  032418 (2018)
	
	\bibitem{andrews2017smoldyn}
	Andrews, S.S.: Smoldyn: particle-based simulation with rule-based modeling,
	improved molecular interaction and a library interface. Bioinformatics
	\textbf{33}(5),  710--717 (2017)
	
	\bibitem{deutsch2005mathematical}
	Deutsch, A., Dormann, S.: Mathematical Modeling of Biological Pattern
	Formation. Springer (2005)
	
	\bibitem{wolf2004lattice}
	Wolf-Gladrow, D.A.: Lattice-gas Cellular Automata and Lattice Boltzmann Models:
	an Introduction. Springer (2004)
	
	\bibitem{wieczorek2008influence}
	Wieczorek, G., Zielenkiewicz, P.: Influence of macromolecular crowding on
	protein-protein association rates—a brownian dynamics study. Biophysical
	Journal  \textbf{95}(11),  5030--5036 (2008)
	
	\bibitem{gardinerStochasticMethodsHandbook2009}
	Gardiner, C.: Stochastic {{Methods}}: {{A Handbook}} for the {{Natural}} and
	{{Social Sciences}}. Springer {{Series}} in {{Synergetics}},
	{Springer-Verlag}, {Berlin Heidelberg}, fourth edn. (2009)
	
	\bibitem{baras1996reaction}
	Baras, F., Mansour, M.M.: Reaction-diffusion master equation: a comparison with
	microscopic simulations. Physical Review E  \textbf{54}(6), ~6139 (1996)
	
	\bibitem{gillespieStochasticSimulationChemical2007}
	Gillespie, D.T.: Stochastic {{simulation}} of {{chemical kinetics}}. Annual
	Review of Physical Chemistry  \textbf{58},  35--55 (2007)
	
	\bibitem{loskotComprehensiveReviewModels2019}
	Loskot, P., Atitey, K., Mihaylova, L.: Comprehensive {{review}} of {{models}}
	and {{methods}} for {{inferences}} in {{bio-chemical reaction networks}}.
	Frontiers in Genetics  \textbf{10}, ~549 (2019)
	
	\bibitem{ocalParameterEstimationBiochemical2019}
	{\"O}cal, K., Grima, R., Sanguinetti, G.: Parameter estimation for biochemical
	reaction networks using {{Wasserstein}} distances. Journal of Physics A:
	Mathematical and Theoretical  \textbf{53}(3),  034002 (2019)
	
	\bibitem{shahriariTakingHumanOut2016}
	Shahriari, B., Swersky, K., Wang, Z., Adams, R.P., {de Freitas}, N.: Taking the
	{{human out}} of the {{loop}}: {{a review}} of {{bayesian optimization}}.
	Proceedings of the IEEE  \textbf{104}(1),  148--175 (2016)
	
	\bibitem{brochuTutorialBayesianOptimization2010}
	Brochu, E., Cora, V.M., {de Freitas}, N.: A {{tutorial}} on {{bayesian
			optimization}} of {{expensive cost functions}}, with {{application}} to
	{{active user modeling}} and {{hierarchical reinforcement learning}}.
	arXiv:1012.2599 [cs]  (2010)
	
	\bibitem{villaniOptimalTransportOld2009}
	Villani, C.: Optimal {{Transport}}: {{Old}} and {{New}}. Grundlehren Der
	Mathematischen {{Wissenschaften}}, {Springer-Verlag}, {Berlin Heidelberg}
	(2009)
	
	\bibitem{rasmussenGaussianProcessesMachine2006}
	Rasmussen, C.E., Williams, C.K.I.: Gaussian Processes for Machine Learning.
	Adaptive Computation and Machine Learning, {MIT Press}, {Cambridge, Mass}
	(2006)
	
	\bibitem{vazquez2010convergence}
	Vazquez, E., Bect, J.: Convergence properties of the expected improvement
	algorithm with fixed mean and covariance functions. Journal of Statistical
	Planning and Inference  \textbf{140}(11),  3088--3095 (2010)
	
	\bibitem{allen2017computer}
	Allen, M.P., Tildesley, D.J.: Computer Simulation of Liquids. Oxford University
	Press (2017)
	
	\bibitem{thomasSlowscaleLinearNoise2012}
	Thomas, P., Straube, A.V., Grima, R.: The slow-scale linear noise
	approximation: An accurate, reduced stochastic description of biochemical
	networks under timescale separation conditions. BMC Systems Biology
	\textbf{6}(1), ~39 (2012)
	
	\bibitem{cranmer2020frontier}
	Cranmer, K., Brehmer, J., Louppe, G.: The frontier of simulation-based
	inference. Proceedings of the National Academy of Sciences  \textbf{117}(48),
	30055--30062 (2020)
	
\end{thebibliography}
\end{document}